\documentclass[twocolumn,showpacs,amsmath,amssymb,pra]{revtex4}

\usepackage{graphicx}% Include figure files
\usepackage{dcolumn}% Align table columns on decimal point
\usepackage{bm}% bold math

\begin{document}

\title
{ERRATUM \\
Solution of periodic Poisson's equation and the Hartree-Fock approach
for solids with extended electron states: 
application to linear augmented plane wave method }

\author{A.V. Nikolaev} 
\affiliation{
Department of Physics, University of Antwerp, UIA, 2610 Antwerpen,
Belgium}
\affiliation{
 Institute of Physical Chemistry of RAS, 
Leninskii prospect 31, 117915, Moscow, Russia 
} 

\author{P.N. Dyachkov}
\affiliation{
Kurnakov Institute of General and Inorganic Chemistry of RAS,
Moscow, Russia}

\date{\today} 

%--------------- ABSTRACT --------------- 
\begin{abstract} 
Int. J. Quantum Chem. {\bf 89}, pp. 57-85 (2002)
\end{abstract} 
 
\pacs{31.10.+z, 31.25.-v, 75.75.+a, 73.21.-b} 

\maketitle

We would like to notice that
the issue of the Hartree-Fock LAPW (linear augmented plane wave) approach
was considered before by S. Massidda, M. Posternak, and A. Baldereschi
in Phys. Rev. B {\bf 48}, pp. 5058-5068 (1993), but we were unaware
of this publication.
We thank D.J. Singh and M. Weinert for this reference.

Eq. (2.5a) is missing the contributions from $\vec{K} \neq 0$, i.e.
it should read as 
\begin{eqnarray}
 \rho'_0(r) &=& \rho_0(r)-\sqrt{4 \pi} \rho_I(\vec{K}=0) 
 -Z\, \frac{\delta(r)}{\sqrt{4\pi} r^2} \nonumber \\ 
 & &-\sqrt{4\pi} 
 {\sum_{\vec{K} \neq 0}}' j_0(Kr) \rho_I(\vec{K}) . 
 \quad \quad \quad \quad \quad \quad (2.5a) \notag
\end{eqnarray}
Correspondingly, Eqs.\ (2.11) acquire additional terms:
\begin{mathletters}
\begin{eqnarray}
  Q_0(r) &=& 
 q_0(r)-\frac{4\pi r^3}{3} \rho_I(\vec{K}=0) \nonumber \\
 & & -4\pi r^2 {\sum_{\vec{K} \neq 0}}' 
      \frac{j_1(Kr)}{K} \rho_I(\vec{K}) ,
 \quad \quad \quad \quad \quad (2.11a) \notag
    \\
 Q'_0(r) &=& q'_0(r)
 -2 \pi (R^2-r^2)\, \rho_I(\vec{K}=0) \nonumber \\ 
 +& 4 \pi & {\sum_{\vec{K} \neq 0}}' (\cos(KR)-\cos(Kr)) 
 \frac{\rho_I(\vec{K})}{K^2} .
 \quad \quad  (2.11b) \notag
\end{eqnarray}
\end{mathletters}
The additional terms are the last ones in 
Eq.~(2.5a), Eq.~(2.11a) and Eq.~(2.11b).\\
{\it For the cases which we consider in Sec. 3 though this does not make any
difference}.

Eq.\ (3.3) should read as
\begin{eqnarray}
  E_{fcc}/N=\frac{1}{2} \left( V_0^{out}-
   \langle V _I \rangle \right) =-1.2079 \; \mbox{ a.u.}  
  \quad \quad \quad  (3.3) \nonumber
\end{eqnarray}
More details on the Madelung potential are given by M. Weinert,
E. Wimmer, and A.J. Freeman in Phys. Rev. B {\bf 26}, 4571 (1982).

We thank M. Weinert for these corrections and for discussing the results
of our work.

Finally, for clarity we would like to notice that the exact solutions
$V_{04}(exact)$ and $V_{06}(exact)$ in Figures 2 and 3 are given 
explicitly by
\begin{eqnarray}
   & & V_{04}^{(exact)}(r)=\tilde{V}_{04} \left( \frac{r}{R} \right)^4, \nonumber \\
   & & V_{06}^{(exact)}(r)=\tilde{V}_{06} 
   \left( \frac{r}{R} \right)^6. \nonumber 
\end{eqnarray}
Here $\tilde{V}_{04}=-0.18192$, $\tilde{V}_{06}=-0.14466$ are exact values
at the radius $R=\sqrt{2}/4$
(see last row of Table II). Thus, in Figures 2 and 3 we depict the deviations
from those exact dependences for 
two solutions (Weinert and Ewald) when $r < R$.

\end{document}